\documentclass[aps,twocolumn,nofootinbib,showpacs,prd,aps,10pt]{revtex4}
\usepackage[dvips]{graphicx}
\usepackage[english]{babel}
\usepackage[T1]{fontenc}
\usepackage{mathrsfs}
\usepackage[tbtags]{amsmath}
\usepackage{amssymb}
\usepackage{amsxtra}
\usepackage{amsopn}
\usepackage{latexsym}
\usepackage[mathcal]{eucal}
\usepackage{mathtools}

\newcommand{\BE}{\begin{equation}}
\newcommand{\EE}{\end{equation}}
\newcommand{\BA}{\begin{align}}
\newcommand{\EA}{\end{align}}
\newcommand{\Tr}{\mathrm Tr}
\newcommand{\nn}{\nonumber}
\newcommand{\kkk}{ \frac{{\rm d}^4k}{(2\pi)^4}}
\newcommand{\ppp}{ \frac{{\rm d}^4p}{(2\pi)^4}}
\newcommand{\fsl}[1]{\ensuremath{\mathrlap{\!\not{\phantom{#1}}}#1}}
\begin{document}

\title{Higher order extensions of the Gaussian effective potential}

\author{Fabio Siringo}

\affiliation{Dipartimento di Fisica e Astronomia 
dell'Universit\`a di Catania,\\ 
INFN Sezione di Catania,
Via S.Sofia 64, I-95123 Catania, Italy}

\date{\today}
\begin{abstract}
A variational method is discussed, extending the Gaussian effective potential to higher orders. 
The single variational parameter is replaced by trial unknown two-point functions, 
with infinite variational parameters to be optimized by the solution of a set of integral equations. 
These stationary conditions are derived by the self-energy without having to write the effective potential, 
making use of a general relation between self-energy and functional derivatives of the potential. 
This connection is proven to any order and verified up to second order 
by an explicit calculation for the scalar theory. 
Among several variational strategies, the methods of minimal sensitivity and  
of minimal variance are discussed in some detail.  
For the scalar theory, at variance with other post-Gaussian approaches, 
the pole of the second-order propagator is shown to satisfy the simple first-order gap equation 
that seems to be more robust than expected. 
By the method of minimal variance, nontrivial results are found for gauge theories containing fermions, 
where the first-order Gaussian approximation is known to be useless.
\end{abstract}


\maketitle

\section{introduction}

Variational methods are very useful in quantum mechanics,  and  quite often their use
becomes mandatory when the interaction strength is too large.
In quantum field theory, the use of variational methods has always been questioned because of the
weight of high energy modes that prevents any reasonable physical result unless the trial
functional has the exact high energy asymptotic  behavior.
Recently, a new interest has emerged on variational methods\cite{kogan,reinhardt,szcz} 
because of the relevance of non-Abelian gauge theories that are known to be asymptotically free. 
The high energy asymptotic
behavior of these theories is known exactly, while the low energy physics can only be accessed by
numerical lattice simulations because of the large strength of the interaction that does not
allow the use of standard perturbation theory. 
As a consequence, important problems like quark confinement and the low energy phase diagram
of QCD still lack a consistent analytical description, and the development of nonperturbative 
variational techniques would be more than welcome in this important area of quantum field theory.

Another problem with variational methods is calculability: the energy is a functional
of the quantum fields and, while in principle any trial functional could be chosen, the need
of  an analytically tractable theory makes the Gaussian functional the only viable choice.
Thus, we are left with the Gaussian Effective Potential (GEP), which has been discussed by several
authors\cite{schiff,rosen,barnes,stevenson}  
and, at variance with perturbation theory,  there is no obvious way to improve the approximation
order by order.
 
Besides being a truly variational method, the GEP may also be regarded as a self-consistent
theory since the  proper self-energy vanishes at first order, and the Gaussian free-particle
Green function is equal to the first-order function.
The GEP has many merits and has been successfully applied to physical problems ranging from
electroweak symmetry breaking\cite{ibanez} 
and scalar theories\cite{stevenson,var,light,bubble} in 3+1 spacetime dimensions, to superconductivity
in bulk materials\cite{superc1,superc2} and films\cite{kim}, to non-Abelian gauge theories\cite{su2} 
and quite recently to the Higgs-top sector of the standard model\cite{LR,AF,HT}.

Even if  the GEP usually gives a fairly good representation of reality,  
an extension of the Gaussian approximation has always been desirable.
However, any attempt to improve the GEP has not been so successful, and most merits of the GEP
seem to disappear at second order. For instance, the Post-Gaussian Effective Potential (PGEP) discussed
by Stancu and Stevenson\cite{stancu} is not a truly variational method (the exact effective potential cannot 
be shown to be smaller than the PGEP), it is not self-consistent, and it fails to reach a minimum for any finite value 
of the variational parameters (in most cases the vanishing of a second derivative is required).

In this paper, we point out that most of the shortcomings of the PGEP could be 
just a consequence of using a fixed shape for the two-point Gaussian  correlator.
We explore a more general extension of the GEP,  where the best Gaussian two-point function 
is the solution of a nonlinear integral equation, a generalized stationary condition that replaces 
the simple first-order gap equation. At  any order, the generalized stationary condition can be derived by
the self-energy graphs, without having to write the effective potential. Order by order, 
that is possible because of the existence of a simple exact connection between the gap equation and the self-energy 
that allows for a direct derivation of the generalized gap equation by standard methods of perturbation theory. 
The connection between self-energy and gap equation generalizes  
the well-known property of self-consistency of the first-order GEP, which in turn is a consequence of the 
equivalence between the gap equation and the vanishing of the first-order self-energy. 

For a scalar theory under general physical assumptions and for different choices of variational strategies,
we show that the second-order  two-point function is characterized by a self-consistent mass which is formally 
given by the same first-order Gaussian gap equation. Thus, the first-order gap equation seems 
to be more robust than expected. 

The formalism can be extended to more general theories like gauge theories containing bosons and fermions
and might be used for the development of variational approaches to strongly
interacting  sectors of the standard model. The simple case of a $U(1)$ gauge theory with a single
fermion is discussed in some detail and, at variance with GEP and PGEP that are known to be
useless for fermions\cite{stancu,stancu2}, the present method provides an integral equation
with nontrivial solutions that can be evaluated by iterative numerical techniques.

In Section II, after a brief discussion on the viable variational strategies in field theory, an
extension is presented where the finite set of variational parameters is replaced by a trial function 
that is equivalent to an infinite set of variational parameters. The method is illustrated by the simple
model of a self-interacting scalar theory.

In Section III, the proof is given of a general connection between the functional derivative of 
the effective potential and the self-energy. The connection is shown to be valid order by order and plays
a key role for determining the variational stationary conditions without having to write the effective
potential. Up to second order the relation is verified in detail in the Appendix by a direct evaluation
of the effective potential.

In Section IV, the second order extension of the GEP is described in some detail for the scalar theory.
The methods of minimal sensitivity and of minimal variance are compared and shown to provide nonlinear 
integral equations. 
The analytical properties of the solution are studied and the pole of the second-order propagator 
is shown to satisfy the same first-order gap equation of the GEP.

In Section V, the method is extended to the simple $U(1)$ gauge theory with a single fermion,
and the method of minimal variance is shown to be suited for a second-order variational approach
to gauge theories. The stationary conditions provide a linear equation for the propagator with a
nontrivial unique solution. A perturbative expansion of the result is shown to give 
back the standard equations of quantum electrodynamics.

In Section VI, the results of the paper are discussed with  some concluding remarks.

Details on the derivation of the effective potential up to second order are reported in the Appendix.

\section{Generalizations of the Gaussian effective potential}

The Post-Gaussian effective potential was discussed by  Stancu and Stevenson\cite{stancu} for the scalar theory 
with and without fermions\cite{stancu2}. 
One of the main merits of the method is its use of the standard perturbative
techniques for evaluating the effective potential while retaining a variational nature.
In fact the method consists in a perturbative expansion around a trial zeroth-order two-point function
that is then optimized by the variation of a parameter. Since the original Lagrangian does not depend
on the variational parameter, the principle of minimal sensitivity\cite{minimal} is enforced by requiring 
that the nth-order effective potential should be stationary with respect to the variation of the parameter. 
The resulting expansion turns out to be convergent even when the original 
interaction did not contain any small parameter. 
In the original PGEP the two-point function was taken as a free propagator with the mass 
that played the role of the variational parameter. 

The method can be generalized as follows:  
the zeroth-order two-point function could be taken as a free unknown trial function, 
that is equivalent to deal with an infinite set of variational parameters. 
The variational constraint becomes an integral equation for the unknown trial two-point function, 
and the eventual solution would improve over the PGEP.  We have infinitely more variational parameters
while retaining the Gaussian shape of the functional that allows for calculability. 
Of course, no general proof can be given of the existence of a solution, and the problem has to
be studied case by case. Moreover, several different variational constraints and strategies can be proposed 
in order to extend the method order by order, and the existence of a solution depends on the chosen strategy.  

\subsection{Variational methods and strategies}

Consider a quantum field theory with action $S[\phi]$ depending on a set of quantum fields
$\phi\equiv\{\phi_n\}$, and introduce shifted fields $h_n=\phi_n-\varphi_n$ where $\varphi$ is
a set of constant backgrounds.
We can always split the action as
\BE
S[\phi]=S_0[h]+S_I[h]
\EE
where the interaction term is defined as
\BE
S_I[h]=S[\varphi+h]-S_0[h]
\EE
and $S_0[h]$ is a trial functional that can be freely chosen. We take the $S_0$  quadratic in the fields
$h$ in order to get Gaussian integrals that can be evaluated exactly.  This functional can be thought 
to be the free action of a field theory, and we denote by $H_0$ the Hamiltonian of that theory, while
$H$ is the Hamiltonian of the full interacting theory described by the action $S$.
The effective action $\Gamma[\varphi]$ can be evaluated by perturbation theory order by order as a sum of 
Feynman diagrams according to the general path integral representation
\BE
e^{i\Gamma[\varphi]}= \int_{1PI} {\cal D}_h
e^{iS[\varphi+h]}
=\int_{1PI} {\cal D}_h e^{iS_0[h]} e^{iS_I[h]} 
\label{path}
\EE
that is equivalent to the sum of all the one-particle-irreducible (1PI) vacuum diagrams for
the action functional $S[\varphi+h]$, where $\varphi$ acts like a source\cite{weinbergII},
and $S_I$ plays the role of the interaction. In general, the action terms $S_0$ and $S_I$
have an implicit dependence on $\varphi$ that is omitted for brevity.
Denoting by $\langle X\rangle$ the quantum average
\BE
\langle X\rangle = \frac{ \int_{1PI} {\cal D}_h e^{iS_0[h]} X}
{ \int {\cal D}_h e^{iS_0[h]} }
\EE
the effective action can be written as
\BE
i\Gamma[\varphi]=i\Gamma_0[\varphi] + \log\langle e^{iS_I}\rangle
\EE
where the zeroth-order contribution can be exactly evaluated since $S_0$ is
quadratic
\BE
i\Gamma_0[\varphi]= \log\int {\cal D}_h e^{iS_0[h]}
\label{gamma0}
\EE
and  the remaining terms can be written by expansion of the logarithm in moments of $S_I$,
\begin{align}
\log\langle e^{iS_I }\rangle=\sum_{n=1}^{\infty}  i\Gamma_n[\varphi]&=
\langle iS_I \rangle+\frac{1}{2!} 
\langle [ iS_I-\langle iS_I\rangle]^2 \rangle\nn\\
&+\frac{1}{3!} \langle [ iS_I-\langle iS_I\rangle]^3\rangle+\dots
\label{log}
\end{align}
which is equivalent to taking the sum of all connected  1PI vacuum diagrams arising from the
interaction $S_I$, as emerges from a direct evaluation of the averages by Wick's theorem.
In this paper, we use the convention that $\Gamma_n$, $V_n$, $\Sigma_n$ represent the 
single nth-order contribution, while the sum of terms up to nth order are written as
$\Gamma^{(n)}$,  $V^{(n)}$, $\Sigma^{(n)}$, so that
\BE
i\Gamma ^{(N)}=\sum_{n=0}^{N} i\Gamma_n.
\EE
The effective potential follows as $V(\varphi)=-\Gamma[\varphi]/\Omega$ where $\Omega$
is a total spacetime volume. 

On the other hand, the effective potential is known  to be the vacuum energy density $E$, 
and can be expanded around the ground state $\vert 0\rangle$ of $H_0$ in powers of the
interaction $H-H_0$,
\BE
E=E_0+E_1+E_2+\dots
\EE
where $E_0$ is the exact ground-state energy of $H_0$,
\BE
E_0=\langle 0\vert H_0 \vert 0\rangle
\EE
and $E_1$ is the first-order correction
\BE
E_1=\langle 0\vert H- H_0 \vert 0\rangle.
\EE
By a direct comparison of the expansions we see that $V_n=E_n$, and the sum of the first two terms
must give the first-order approximation for the effective potential
\BE
V^{(1)}=E_0+E_1=\langle 0\vert H \vert 0\rangle
\EE
which is the expectation value of the full Hamiltonian $H$ in the trial state $\vert 0 \rangle$. 
Any variation of the parameters in $S_0$ is equivalent to a variation of $H_0$ and
its ground state $\vert 0\rangle$. Thus, a stationary condition imposed on the first-order
effective potential is equivalent to the standard variational method of quantum mechanics.
The resulting optimized first-order effective potential is the Gaussian effective potential.

Extensions of the GEP are not trivial: the second-order approximation for the effective potential 
gives
\BE
V^{(2)}=\langle 0\vert H \vert 0\rangle+E_2
\label{unbound}
\EE
and a variation of the free parameters in $S_0$ is not equivalent to a variation of the
expectation value of $H$. Moreover, it is well known that the second-order correction  $E_2$ is  
negative for any quantum mechanical system, and $V^{(2)}$ can be lower than the exact vacuum energy. 
Thus the simple search for a minimum
of $V^{(2)}$ would not work. Since the exact action $S$ does not depend on the free parameters
in $S_0$, any extension of the GEP requires a new prescription for determining the free parameters.
There are at least three methods that have been suggested: a fixed variational basis\cite{tedesco},
the minimal sensitivity\cite{minimal}, and the minimal variance\cite{sigma,sigma2}.

(i) The parameters in $S_0$  might  be fixed by the minimal of the first-order effective potential, 
that is a genuine variational method. Then the higher order contributions could be evaluated by perturbation
theory with the parameters kept fixed, even if that would spoil the convergence of the expansion. 
The ground state $\vert 0\rangle$ and the other eigenstates of $H_0$ are then used as a fixed
basis set optimized by the first-order variational method, as shown in Ref.\cite{tedesco}.

(ii) Since the exact effective potential does not depend on the variational parameters in
$S_0$, the minimal sensitivity of $V^{(n)}$ as been proposed as a variational criterion\cite{minimal}. 
At each order the parameters are fixed by the stationary point of the total effective potential 
(or its derivative when no solution occurs). The stationary condition changes order by order, 
and the parameters must be determined again at any order. This procedure has been proven to improve
the convergence of the expansion\cite{minimal}.   

(iii) More recently\cite{sigma,sigma2} the search for the minimal variance has been shown to be a valuable 
variational criterion for determining the unknown parameters.  
It is based on the physical idea that in the exact eigenstates of an operator ${\cal O}$, the
variance must be zero because $\langle {\cal O}{\cal O}\rangle=\langle {\cal O}\rangle^2$.
For any Hermitian operator like $H$, the variance is a positive quantity, bounded
from below, and the variational parameters can be tuned by requiring that the variance is minimal.

In quantum mechanics, the last method is not very useful because the accuracy of the standard variational 
approximation can be easily improved by a better trial wave function with more parameters.
In field theory, calculability does not leave too much freedom 
in the choice of the wave functional that must be Gaussian. 
When the simple stationary condition fails, a second-order extension  can be achieved 
by the method of minimal variance, as discussed in Ref.\cite{sigma}.
Actually, we can write the second-order contribution to the effective potential in Eq.(\ref{log}) as
\BE
V_2=E_2=-\frac{\sigma_I^2}{2\Omega}
\label{V2}
\EE
where $\sigma_I$ is the variance of the Euclidean action $S_I^E$,
\BE
\sigma_I^2=\langle (S^E_I)\rangle^2-\langle (S^E_I)^2 \rangle.
\EE
That  follows immediately from Eq.(\ref{log}) by Wick rotating as
the operator $(iS)$ becomes the Euclidean action $(iS)\to -S^E$, while the quantum action 
$i\Gamma\to -V/\Omega$. Eq.(\ref{V2}) is  in agreement with the general requirement that $E_2<0$.
The variance would be zero if $\vert 0\rangle$ were an exact eigenstate of $S_I$, while a
minimal variance is expected to optimize the convergence of the expansion. The free parameters
can be fixed by a stationary condition for the second-order term of the
effective potential $V_2$. Then the optimized variational basis can be used for the evaluation of
the higher order correction, as for the method of Ref.\cite{tedesco}. At variance with that method,
the second-order correction is used instead of the first-order one, and that would be useful whenever
the simple  first-order method should fail, as in gauge theories.

\subsection{Generalization to infinite parameters}

The higher order extensions of the GEP can be generalized to the case of infinite variational parameters. 
The method is illustrated in this section by the simple model of a self-interacting scalar theory.
The Lagrangian reads
 \BE
{\cal L}=\frac{1}{2}\partial^\mu\phi\>\partial_\mu\phi
-\frac{1}{2} m_B^2\phi^2-\lambda\phi^4
\label{Lbare}
\EE
In the spirit of background field method let us introduce a shifted field
$h=\phi-\varphi$ where $\varphi$ is a constant background. The action functional
$S[\phi]$ can be written as

\begin{align}
S[\varphi+h]=&\frac{1}{2}\int h(x) g_B^{-1} (x,y) h(y) {\rm d}^4x{\rm d}^4y \nn \\
&+\int \left( \frac{1}{2} m_B^2 h^2 -V_c [\varphi+h]\right) {\rm d}^4x
\label{shift}
\end{align}
where $V_c$ is the classical potential
\BE
V_c[\phi]=\frac{1}{2} m_B^2\phi^2+\lambda\phi^4
\EE
while $g_B^{-1}(x,y)$ is the bare inverse propagator
\BE
g_B^{-1}(x,y)=(-\partial_\mu\partial^\mu-m_B^2)\delta^4(x-y).
\label{gbare}
\EE

Let us denote by $g(x,y)$ a trial unknown two-point function, and write
the action functional as 
\BE
S[\varphi+h]=S_0[h]+S_I[h]
\EE
where $S_0$ plays the role of the zeroth-order
action functional
\BE
S_0[h]=\frac{1}{2}\int h(x) g^{-1} (x,y) h(y) {\rm d}^4x{\rm d}^4y
\EE
and $S_I$ is the interaction term
\begin{align}
S_I[h]&=\frac{1}{2}\int h(x) \left[g_B^{-1} (x,y)-g^{-1} (x,y)\right] h(y) {\rm d}^4x{\rm d}^4y\nn\\
&+\int \left(\frac{1}{2} m_B^2 h^2 -V_c [\varphi+h]\right) {\rm d}^4x.
\label{SI}
\end{align}
An implicit dependence on $\varphi$ is assumed in $g$, $S_0$, and $S_I$.
Of course, the trial function $g^{-1}$ cancels in the total action $S[\varphi+h]$, which
is exact and cannot depend on it. Thus, this formal decomposition holds for any arbitrary
choice of the trial function, provided that the integrals converge.

The effective action $\Gamma[\varphi]$ can be evaluated by perturbation theory order by order 
as a sum of Feynman diagrams according to the general path integral representation of Eq.(\ref{path})
that is equivalent to the sum of all 1PI vacuum diagrams for
the action functional $S[\varphi+h]$, where $\varphi$ acts like a source.
According to our decomposition of the action functional, we must associate the trial propagator
$g(x,y)$ to the free-particle lines of the diagrams, while the vertices are read from
the interaction terms in $S_I$. The effective action follows order by order as the sum of  connected
diagrams according to  Eqs.(\ref{gamma0}) and (\ref{log}). 
The zeroth-order contribution follows from Eq.(\ref{gamma0}),
\BE
V_0(\varphi)=-\frac{i}{2}\int \frac{{\rm d}^4 k}{(2\pi)^4} 
\log\left[\frac{i}{2\pi} g^{-1}(k)\right]
\label{V0}
\EE
where a  Fourier transform $g^{-1}(k)$ has been introduced for the trial function 
$g^{-1}(x,y)\equiv g^{-1}(x-y)$, while according to Eq.(\ref{gbare}) the bare propagator 
reads $g_B^{-1}(k)=k^2-m_B^2$.

Higher order terms follow by Eq.(\ref{log}) and can be described by standard Feynman diagrams
in terms of connected 1PI graphs. The vertices are extracted from the interaction Eq.(\ref{SI})
that yields the interaction Lagrangian
\begin{align}
{\cal L}_{int}=v_0+v_1h(x)&+v_3h^3(x)+v_4h^4(x)\nn\\
&+\int h(x) v_2(x,y) h(y){\rm d}^4y
\label{Li}
\end{align}
where
\begin{align}
v_0&=-V_c(\varphi)\nn\\
v_1&=-\varphi m_B^2-4\lambda\varphi^3\nn\\
v_2&=\frac{1}{2}(g_M^{-1}-g^{-1})\nn\\
v_3&=-4\lambda\varphi\nn\\
v_4&=-\lambda
\label{vi}
\end{align}
and the modified bare propagator $g_M^{-1}(k)=k^2-M^2$ is defined in terms
of the shifted mass
\BE
M^2=m_B^2+12\lambda\varphi^2.
\EE

Up to second order, the connected 1PI vacuum diagrams are shown in Fig.1 with their
symmetry factors.
The first-order contribution  is given by the sum of the tree, one-loop,  and two-loop
graphs in the first row of Fig.1,
\begin{align}
i\Gamma_1[\varphi]&=i\Omega v_0+\int i v_2(x,y) i g(y,x) {\rm d}^4x{\rm d}^4y\nn\\
&+3iv_4\int ig(x,x)ig(x,x)  {\rm d}^4x.
\end{align}

\begin{figure}[b] \label{fig:graphs}
\centering
\includegraphics[width=0.32\textwidth,angle=-90]{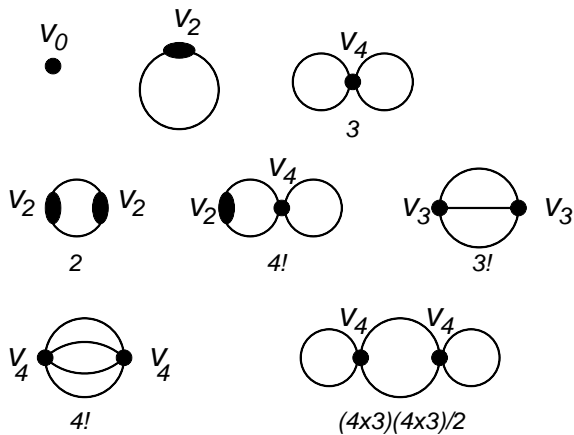}
\caption{Connected vacuum 1PI graphs  for the Lagrangian Eq.(\ref{Li}), up to second order.
The trial propagator $g$ is reported as a straight line, while the vertices are defined in
Eq.(\ref{vi}). The symmetry factors are displayed below the graphs.}
\end{figure}

Neglecting a constant term and dividing by a spacetime volume $\Omega$, the  first-order
contribution to the effective potential reads
\begin{align}
V_1(\varphi)&=V_c(\varphi)-\frac{i}{2}\int\kkk g^{-1}_M(k)g(k)\nn\\
&+3\lambda\left[\int \kkk ig(k)\right]^2
\end{align}
Adding the zeroth-order term of Eq.(\ref{V0}), the first-order effective potential can be written as
\BE
V^{(1)}(\varphi)=V_c(\varphi)+I_1[g]+3\lambda (I_0[g])^2-\frac{i}{2}\int\kkk g^{-1}_M(k)g(k)
\label{V1}
\EE
where the functionals $I_n[g]$ have been defined as a generalization of the GEP and PGEP notation
of Ref.\cite{stancu}:
\begin{align}
I_0[g]&=\int \kkk ig(k)\nn\\
I_1[g]&=-\frac{i}{2}\int \kkk \log\left[\frac{i}{2\pi} g^{-1}(k)\right]\nn\\
I^{(n)}[g]&=n!i\int [ig(x,y)]^n {\rm d}^4x\>{\rm d}^4y.
\label{I}
\end{align}
All the divergent integrals are supposed to be regularized by a cutoff or other regularization scheme.

While the technique is based on perturbation theory, the approximation is valid even when 
there are no small parameters, provided that the effective potential is optimized by a variational criterion.
In fact the interaction is defined in terms of the unknown trial function $g$, and its variation
has an effect on both $S_0$ and $S_I$. The optimal choice of this pair should be the one  that makes
the effects of the interaction $S_I$ smaller in the vacuum of $S_0$. 
A stationary condition can be imposed by requiring that the
functional derivative  is zero. For instance,  the principle of minimal sensitivity would require
that, at a given order, the trial function $g$ satisfies the stationary condition
\BE
\frac{\delta V^{(n)}}{\delta g}=0.
\label{delVn}
\EE
In general this is a nonlinear integral equation for the unknown function $g$. Unfortunately we have
no general proof of the existence of a solution, and the problem should be studied order by order.

At first order we require that
\BE
0=\frac{\delta V^{(1)}}{\delta g(k)}=\frac{i}{2}\left(g^{-1}(k)-g_M^{-1}(k)+12\lambda I_0[g]\right)
\label{delV1}
\EE
and find the simple solution
\BE
g^{-1}(k)=k^2-M_1^2
\label{g1}
\EE
where the mass $M_1$ is the solution of the first-order gap equation
\BE
M_1^2=m_B^2+12\lambda\varphi^2+12\lambda I_0[g]
\label{M1}
\EE
By inserting the self-consistent solution of Eqs.(\ref{M1}) and (\ref{g1})  in  Eq.(\ref{V1}), we
obtain the standard GEP\cite{stevenson,stancu}. Thus, at first order, the present method is equivalent to the GEP. 
The extra freedom on the shape of $g$ does not add anything to the
first-order approximation, and the best $g$ maintains the form of a free-particle propagator.
On the other hand, the shape of the optimized
propagator tells us that the wave function renormalization is negligible in  the self-interacting scalar theory, 
as confirmed by several lattice calculations. Moreover we can show that the first-order proper self-energy vanishes, 
so that the first-order propagator is equal to the optimized function
$g$ that can be regarded as a self-consistent solution at first order. 
The graphs contributing to the self-energy are reported in Fig.2  
up to second order (tadpole graphs are not included). 
The proper first-order term is the sum of the tree and the one-loop graphs in the first row,

\begin{figure}[t] \label{fig:self}
\centering
\includegraphics[width=0.45\textwidth,angle=0]{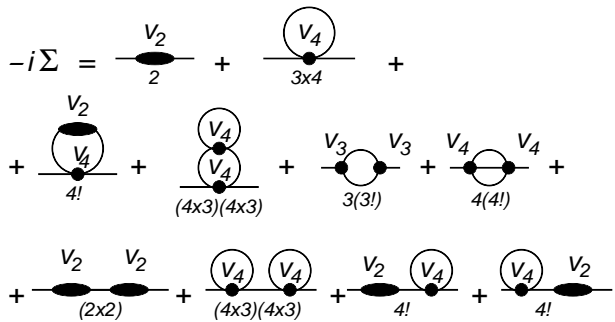}
\caption{Self-energy graphs  for the Lagrangian Eq.(\ref{Li}), up to second order.
The trial propagator $g$ is reported as a straight line, while the vertices are defined in
Eq.(\ref{vi}). The symmetry factors are displayed below the graphs. We recognize two first-order
graphs in the first row, four 1PI graphs in the second row (proper self-energy), and four
reducible graphs in the last row. Tadpole graphs are not included.}
\end{figure}

\BE
-i\Sigma_1=i(g^{-1}_M-g^{-1})+12iv_4 \int \kkk ig(k)
\EE
yielding
\BE
\Sigma_1(k)=g^{-1}(k)-k^2+m_B^2+12\lambda\varphi^2+12\lambda I_0[g]
\label{sigma1}
\EE
which vanishes if the  function $g$ in Eq.(\ref{g1}) satisfies the first-order gap equation Eq.(\ref{M1}).
Actually, by inspection of Eq.(\ref{delV1}), we can see that
\BE
\frac{\delta V^{(1)}}{\delta g(k)}=\frac{i}{2}\Sigma_1.
\label{delV1s}
\EE
This consistency relation is a special case of a more general
relation between self-energy and functional derivatives of the effective potential.
In the next sections we will generalize the result to any order  and   to theories
containing fermions.

For a second-order extension of the method, the trial function $g$ could be determined by
the method of minimal sensitivity as the solution of the stationary condition
\BE
\frac{\delta V^{(2)}}{\delta g}=\frac{\delta}{\delta g}(V_0+V_1+V_2)=0
\label{delV2a}
\EE
or by the method of minimal variance as the solution of the stationary condition
\BE
\frac{\delta V_2}{\delta g}=0
\label{delV2b}
\EE
These are integral equations for the unknown function $g$, and their solution is equivalent
to the optimization of infinite parameters.
In both cases, we would need the functional derivative of the second-order term $V_2$. However,
by a generalization of Eq.(\ref{delV1s}),  
the higher order stationary conditions can be derived  through a simpler path that makes use of the self-energy, 
without having to write the effective
potential.

\section{Connection to self-energy}

It is useful to develop a  general method for the direct evaluation of the functional derivatives
that appear in most of the variational approaches.
We give proof of a general relation between self-energy and functional derivatives of the effective potential. 
For the sake of simplicity, we discuss the case of the self-interacting scalar theory, while
the extension to more complex theories containing Bose and Fermi fields is quite straightforward. An extension
to fermions is discussed below in Section V.
 
At each order, we find that
\BE
\frac{\delta V_n}{\delta g(k)}=\frac{i}{2} \left( \Sigma_n (k)-\Sigma_{n-1}(k)\right)
\label{delVns}
\EE
where $V_n$ and $\Sigma_n$ are the nth-order contribution to the effective potential
and to the self-energy respectively, and no tadpole graph has been included in the self-energy
(as is usually the case at the minimum of $V$ where $\delta V/\delta\varphi=0$ and the tadpoles 
cancel exactly). The relation can be taken to be valid even for $n=1,0$, 
provided that we define $\Sigma_{-1}=0$ and 
$i\Sigma_0=2\delta V_0/\delta g$.
As a corollary, we find that the total nth-order effective potential $V^{(n)}$ satisfies
\BE
\frac{\delta V^{(n)}}{\delta g(k)}=\frac{i}{2}  \Sigma_n(k)
\label{delVsigma}
\EE
and the vanishing of the functional derivative is equivalent to the vanishing of the nth-order
contribution to the self-energy. In the special case of $n=1$, we recover Eq.(\ref{delV1s}) which
 is equivalent to the vanishing of the total self-energy 
and to the self-consistency of the optimized $g$ for the GEP, 
as discussed at
the end of the previous section.

\begin{figure}[b] \label{fig:delta}
\centering
\includegraphics[width=0.25\textwidth,angle=-90]{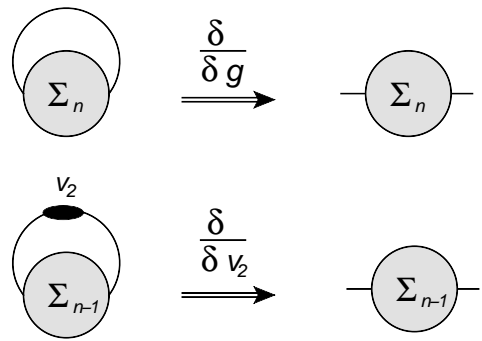}
\caption{Correspondence between vacuum and self-energy graphs: in the upper row,
vacuum 1PI connected graphs of order $n$ give connected self-energy graphs of
order $n$ (without tadpoles) when operating with a  partial functional derivative $\delta/\delta g$
according to Eq.(\ref{delVpartial}); in the lower row, 
vacuum 1PI connected graphs of order $n$ give connected self-energy graphs of order $n-1$ (without tadpoles) 
when operating with a  functional derivative of the vertex  $v_2$.}
\end{figure}

The proof follows by Wick's theorem and inspection of the diagrams. First of all, let us recall a general
relation between the vacuum diagrams without any external line and the self-energy two-point
diagrams with two external vertices where two external lines can be attached. 
All diagrams contributing to the nth order self-energy can be drawn by taking n interaction terms, 
picking up a pair of fields, and contracting all other fields according to Wick's theorem. 
There is a contribution for each
chosen  pair of external fields. If we contract the external fields, we close the two-point diagram by a line
and obtain a vacuum nth-order diagram without external lines and vertices, as shown in Fig.3. 
All vacuum diagrams can be drawn by picking up a pair of fields in all the possible ways, writing all the
corresponding two-point diagrams, and then closing them by a line. As a consequence of Wick's theorem, 
this procedure ensures that we find the correct symmetry factors. The argument can be reversed, and
provided that we inserted the correct symmetry factors, if we cut a line in any possible
way in all the nth-order vacuum diagrams, we obtain all the nth-order two-point diagrams
contributing to the self-energy. Actually, an overall 2 factor must be added  because of the permutation 
of the external fields in the two-point function. Using Feynman rules in momentum space,
for any internal particle line, a factor $ig(k)$ is  included and integrated over $k$. The functional
derivative $\delta /\delta(ig)$ deletes a factor $ig$ and its corresponding integration; thus, it is 
equivalent to the cut of an internal line in all the possible ways, with the correct factor coming out
from the derivative. Thus, denoting by  $-i\Omega {\cal V}_n$ and  $-i\Sigma_n$ the sum of all the
nth-order  vacuum and two-point graphs, respectively,   
\BE
\left(\frac{\delta {\cal V}_n}{\delta g(k)}\right)_{\displaystyle{v_2}}=\frac{i}{2}  \Sigma_n (k).
\label{delVpartial}
\EE
Here by $\delta /\delta g$ we mean the explicit partial derivative with the vertex $v_2$ kept fixed,
while a total functional derivative would operate on the vertex $v_2$ also. In vacuum diagrams,
the nonlocal two-point vertex $v_2$  is only present inside loops  
that have the following general form:
\BE
\int \kkk\dots ig(k)\cdot i v_2(k)\cdot ig(k)\dots
\EE
and by insertion of the vertex definition  Eq.(\ref{vi}), we see that for each of them the total functional 
derivative acquires the extra term
\begin{align}
&\frac{\delta}{\delta g(p)}\left[ \kkk \dots ig(k) \cdot i v_2(k)\cdot ig(k)\dots\right]=\nn\\
&=\left[\dots ig(p) i\frac{\partial v_2(p)}{\partial g(p)}ig(p)\dots\right]
=\left[\dots \left(-\frac{i}{2}\right)\dots\right]
\label{delvert}
\end{align}
where the functional derivative only acts on the vertex. 
The effect of the derivative is the opening of the loop where the vertex was, 
yielding a two-point graph with a vertex less, as shown in Fig.3.
At a given order, all the $v_2$ insertions in vacuum diagrams are in one-to-one correspondence with 
all the possible pairs of external fields that can be picked up for drawing two-point diagrams.
Again, provided that the correct symmetry factors were included, 
Wick's theorem ensures that the functional derivative of all $v_2$ insertions in ${\cal V}_n$
yields the sum of all two-point diagrams of order $n-1$, with the correct factor that comes out
from the derivative. Thus, inserting the factor $-i/2$ coming out from the derivative of the
vertex in Eq.(\ref{delvert}), and adding the result of the partial derivative Eq.(\ref{delVpartial}),
the total functional derivative is
\BE
\frac{\delta {\cal V}_n}{\delta g(k)}=\frac{i}{2}  \left[\Sigma_n (k)-\Sigma_{n-1}(k)\right].
\label{delV}
\EE
Of course, here ${\cal V}_n$ and $\Sigma_n$ contain all kinds of terms including
disconnected diagrams and tadpoles.

It is not difficult to understand that in Eq.(\ref{delV}), disconnected diagrams for $\Sigma_n$
are in correspondence with disconnected or reducible diagrams for ${\cal V}_n$, while self-energy
diagrams containing tadpoles can only generate reducible vacuum diagrams. On the other hand, 
connected self-energy diagrams without tadpoles always generate 1PI connected vacuum diagrams
when the two-point graph is closed with a line or a  $v_2$ vertex, and 1PI connected vacuum diagrams
always generate connected self-energy diagrams without tadpoles when a line or a $v_2$ vertex is
cut by the functional derivative.  When restricting to 1PI connected vacuum diagrams, ${\cal V}_n$
becomes the nth-order contribution to the effective potential $V_n$, and then Eq.(\ref{delVns})
holds, provided that $\Sigma_n$ contains all connected self-energy graphs without tadpoles.

\section{Second-order extensions of GEP}

The first-order stationary condition Eq.(\ref{delV1}) has been shown to be equivalent 
to the gap equation of the GEP Eq.(\ref{M1}) yielding the simple free-particle propagator of Eq.(\ref{g1}).
As discussed in Section II, the second-order extension of the GEP is not trivial. Here, we study in more
detail the second-order stationary conditions that emerge by the methods of minimal sensitivity 
Eq.(\ref{delV2a}) and minimal variance Eq.(\ref{delV2b}), for the self-interacting scalar theory.
The stationary conditions are derived by the self-energy according to Eq.(\ref{delVns}).

\subsection{Stationary conditions}

By the method of minimal sensitivity, the generalized second-order stationary condition
Eq.(\ref{delV2a})  reads
\BE
\frac{\delta V^{(2)}}{\delta g(k)}=\frac{i}{2}  \Sigma_2 (k) =0
\label{delV2aa}
\EE
where Eq.(\ref{delVsigma}) has been used with $n=2$. Thus, the second-order gap equation is 
equivalent to the vanishing of the second-order contribution to the self-energy, ignoring tadpoles.
All the second-order self-energy graphs are displayed in Fig.2. Adding together the four  1PI graphs in
the second line, the proper second-order self-energy can be written as
\BE
\Sigma^\star_2 (k)=-12v_4I_{\Sigma 1}+v_3^2 J_3(k)+v^2_4J_4(k).
\EE
where $I_{\Sigma1}$, $J_3(k)$, and $J_4(k)$ are new functionals of $g$.
The functional $I_{\Sigma 1}$ does not depend on $k$ and is defined as
\BE
I_{\Sigma 1}= \int \kkk i [g(k)]^2\Sigma_1(k).
\EE
It generalizes the functional $I_0$ by the inclusion of a self-energy insertion in the loop, and
arises from the sum of the first pair of second-order graphs in Fig.2.
The functionals $J_n (k)$ carry an explicit dependence on $k$ and are defined as
\begin{align}
J_n(k)&=n\cdot n! \int g(k+\sum_{j=1}^{n-2} k_j)
\prod_{j=1}^{n-2} \left[ i g(k_j) { \frac{{\rm d}^4k_j}{(2\pi)^4}}\right]=\nn\\
&=\frac{i \delta I^{(n)}}{\delta g(k)} 
\label{Jn}
\end{align}
where the last equality follows by evaluating  the integrals  
$I^{(n)}$ in Eq.(\ref{I}) in momentum space. Here the functionals $J_3$ and $J_4$ arise
from the third and fourth 1PI graphs respectively, as displayed in Fig.2.
The four reducible second-order graphs can be easily expressed in terms of $\Sigma_1$
yielding for the total second-order self-energy
\BE
\Sigma_2 (k)=\Sigma_2^\star (k)+ g(k)[\Sigma_1(k)]^2.
\EE
The  condition of minimal sensitivity  Eq.(\ref{delV2aa}) then reads
\BE
\Sigma_2^\star (k)=- g(k)[\Sigma_1(k)]^2.
\label{sta2}
\EE
The  derivation of this second-order stationary condition was made quite simple
by the use of self-energy graphs and their connection to the functional derivatives
of the effective potential. However the same equation could be derived by the more
cumbersome calculation of  the vacuum diagrams in Fig.1, followed by the functional
derivative of all terms. Since it is instructive to examine the connection between the two methods,  
in the Appendix the effective potential is evaluated up to second order and its functional
derivative is compared with the self-energy, term by term, showing a perfect agreement
with Eq.(\ref{delV2aa}).
 
By the method of minimal variance, the generalized second-order stationary condition 
Eq.(\ref{delV2b}) reads
\BE
\frac{\delta V_2}{\delta g(k)}=\frac{i}{2}( \Sigma_2 (k)-\Sigma_1(k)) =0
\label{delV2bb}
\EE
where Eq.(\ref{delVns}) has been used with $n=2$. That is equivalent to imposing
$\Sigma_2=\Sigma_1$.
In terms of proper self-energies the  condition of minimal variance can be written as
\BE
\Sigma_2^\star (k)=- g(k)[\Sigma_1(k)][\Sigma_1(k)-g^{-1}(k)]
\label{sta2b}
\EE
and differs from the condition of minimal sensitivity Eq.(\ref{sta2}) for the term $g^{-1}$ in the last factor.

Despite the simple shape of the resulting stationary conditions Eqs.(\ref{sta2}),(\ref{sta2b}),
these are nonlinear integral equations for $g$, and we cannot even  prove the existence of a solution.  
It would be interesting to look for a numerical solution, but that is out of the aim of the present paper.

In the PGEP of Ref.\cite{stancu} the trial function $g$ is forced to be the same 
as  for the first-order approximation, but with a mass $M_2$ that should satisfy 
a second-order gap equation coming out from the condition of minimal sensitivity. 
Actually, they find no solution for their single variational parameter $M_2$ 
(their second-order effective potential is never stationary). That does not mean
that the integral equation Eq.(\ref{sta2}) has no solution, since the trial function $g$ has 
infinite degrees of freedom in our generalized approach. On the other hand, with the same constraint of
a free-particle $g$, with a single variational mass parameter, the method of minimal variance has been
shown to give a solution\cite{sigma}. In fact the variance is bounded, and it is more likely to
have a minimum compared to $V^{(2)}$ that could even be unbounded 
according to Eq.(\ref{unbound}). On the same footing, that does not mean that the integral 
equation (\ref{sta2b}) must have a solution, but it is a good physical argument for its existence.

\subsection{Analytical properties of solutions}

Even without having derived the effective potential, some interesting consequences
of the stationary equations can be studied. In fact, the present method allows for a study of 
the existence and properties of the solution without having to write down the effective potential.

Let us suppose that a function $g$ does exist, satisfying one of the stationary 
conditions Eqs.(\ref{sta2}) or (\ref{sta2b}),
and let us look at the analytical properties of this function. We can prove that the single-particle
pole of the function $g$ must be at $k=M_1$, where $M_1$ is the solution of the first-order
gap equation (\ref{M1}). That does not mean that the pole does not change, 
because the function $g$ in Eq.(\ref{M1}) must be the solution of the second-order stationary equation instead
of the simple first-order solution Eq.(\ref{g1}). In other words, the second-order extension does
not change the Gaussian gap equation but changes the shape of the function $g$ from its
first-order free-particle form. That would also explain why no solution is found for the stationary
condition in the PGEP of Ref.\cite{stancu} where a free-particle trial function is used. 
The lack of any solution could be the sign of having chosen a wrong trial function.
On the other hand, the first-order gap equation (\ref{M1}) seems to be more 
robust than expected by the PGEP analysis.

The proof of the above statement comes from 
a more careful inspection of Eqs.(\ref{sta2}) and (\ref{sta2b}).
Let us suppose that a solution $g(k)$ does exist, and that its single-particle pole is at $k=m$. In
other words we are assuming that $k=m$ is the first singular point on the real axis, while other
singularities or cuts may occur for $k>m$. We are going to prove that $m=M_1$, as defined in
Eq.(\ref{M1}).
 
If Eqs.(\ref{sta2}) or (\ref{sta2b})  hold, and $g$ has a pole in $k=m$, 
then the proper self-energy
$\Sigma_2^\star (k)$ must have a pole unless $\Sigma_1$ vanishes.  In this special
point $k=m$,  the two stationary conditions Eqs.(\ref{sta2}) and (\ref{sta2b})  
are equivalent because $g^{-1}(m)=0$.
In $\Sigma^\star_2 (k)$, the dependence on $k$ comes from the functionals $J_3 (k)$ and
$J_4(k)$. These are functionals of $g$ and cannot have a pole for $k<2m$ and $k<3m$,
respectively. That is obvious in Euclidean space where the pole $g$ becomes imaginary and gives
the long-range behavior of the Fourier transform $g(x)\sim exp(-mx)$. The functional $J_{n+1}(k)$
is the Fourier transform of $[g(x)]^{n}\sim exp(-n\cdot mx)$  and cannot have a pole for 
$k<n\cdot m$. That is a well-known property of graphs with multiparticle intermediate states
like the one-loop and two-loop sunrise graphs that give the $J_3$ and $J_4$ terms.
Then the function  $\Sigma_1 (k)$ must have a first-order zero in $k=m$ at least, supposing it
to be a generally analytical function of $k$. As a by-product, $\Sigma^\star_2$ must have
a zero in $k=m$ as well, in order to satisfy Eqs.(\ref{sta2}) or (\ref{sta2b}). 
Then, inserting $\Sigma_1(m)=0$ and
$g^{-1}(m)=0$ in Eq.(\ref{sigma1}), we obtain
\BE
m^2=m_B^2+12\lambda\varphi^2+12\lambda I_0[g]
\label{gap2}
\EE
which means that  $m$ satisfies the first-order gap equation, i.e. $m=M_1$.

Denoting by $G_{(n)}$ the nth-order propagator, as obtained by standard perturbation theory with
the interaction ${\cal L}_{int}$ of Eq.(\ref{Li}), we saw that $G_{(1)}=g$ at first order, i.e. the
first-order approximation is self-consistent since $\Sigma_1$ vanishes identically.
Setting $\varphi$ at its physical value, at the minimum of the effective potential
where $\delta V/\delta\varphi=0$, all tadpole graphs cancel exactly in the self-energy, and
the second-order propagator can be written in terms of proper self-energy insertions as
\BE
G_{(2)} ^{-1}=g^{-1}-\Sigma_1-\Sigma_2^\star
\EE
and inserting Eq.(\ref{sigma1}) we obtain
\BE
G_{(2)} ^{-1}(k)=k^2-M_1^2-\Sigma_2^\star (k)
\EE
which still has a pole in $k=M_1$ since $\Sigma_2^\star(M_1)=0$.
Thus, the pole is still self-consistent even if  $g$  is not, since $g(k)\not= G_{(2)}(k)$ in
the second-order approximation, and a non-vanishing wave function renormalization can
be extracted by the residue of the pole.

While the present results are suggestive, they are all based on the hypothesis that without other
constraints, at least one of the stationary conditions Eqs.(\ref{sta2}) or (\ref{sta2b})
might have a solution. 
Moreover, we have not addressed the issue of renormalization: most of the integrals are divergent and, 
while a simple cut-off regularization would be enough in the case of an effective theory, 
renormalization of the bare parameters would be an interesting aspect to be studied. 
Again, perturbative techniques
can be used as shown in Ref.\cite{stancu}, even if the result has a genuine variational nature.

\section{Gauge interacting Fermions}

For fermions, variational methods like GEP and  PGEP   are known to be useless\cite{stancu2}, as
the methods just reproduce the known results of perturbation theory. Thus, gauge theories with
interacting fermions seem to be an interesting test for the generalized higher order extension of
the GEP. The failure of the first-order GEP is a simple consequence of the minimal 
interaction that in gauge theories does not admit  any first-order vacuum graph. 
It is mandatory to use higher order
approximations, and the method of minimal variance seems to be suited to the case.

Quantum electrodynamics (QED) is the simplest $U(1)$ theory of interacting  fermions. Let us consider
the basic theory of a single massive fermion interacting through an Abelian gauge field 
\BE
{\cal L}=\bar\Psi(i\fsl{\partial}+e  {\ensuremath{\mathrlap{\>\not{\phantom{A}}}A}}
-m)\Psi-\frac{1}{4}F^{\mu\nu}F_{\mu\nu}
-\frac{1}{2}(\partial_\mu A^\mu)^2    
\label{Le}
\EE
where the last term is the gauge fixing term in the Feynman gauge, and the electromagnetic tensor is
$F_{\mu\nu}=\partial_\mu A_\nu-\partial_\nu A_\mu$.

Introducing a shift $a^\mu$ for the gauge field $A^\mu\to A^\mu+a^\mu$, the quantum effective action
follows as the sum of connected  vacuum 1PI graphs that are summarized by the path integral representation
\BE
e^{i\Gamma[a]}=\int_{1PI} {\cal D}_{A}{\cal D}_{\bar\Psi,\Psi}
e^{iS[a+A]}
\EE
where the action $S$ can be split as $S=S_0+S_I$.
We define the trial action $S_0$ as
\begin{align}
S_0&=\frac{1}{2}\int A^\mu(x) D^{-1}_{\mu\nu}(x,y) A^\nu(y) {\rm d}^4x{\rm d}^4y \nn \\
&+\int \bar\Psi(x) G^{-1}(x,y) \Psi (y) {\rm d}^4x{\rm d}^4y
\end{align}
where $D_{\mu\nu}(x,y)$ and $G(x,y)$ are unknown trial matrix functions.
The interaction contains three terms
\begin{align}
S_I&=\frac{1}{2}\int A^\mu(x) 
\left[\Delta^{-1}_{\mu\nu}(x,y)-D^{-1}_{\mu\nu}(x,y)\right] A^\nu(y){\rm d}^4x{\rm d}^4y \nn \\
&+\int \bar\Psi(x)
\left[ g_m^{-1}(x,y)-G^{-1}(x,y)\right] \Psi (y) {\rm d}^4x{\rm d}^4y\nn\\
&+e\int\bar\Psi(x) \gamma^\mu A_\mu(x) \Psi(x){\rm d}^4x
\label{SI2}
\end{align}
where $\Delta_{\mu\nu} (x,y)$ and $g_m (x,y)$ are free-particle propagators. Their
Fourier transform can be expressed as
\begin{align}
\Delta_{\mu\nu}^{-1} (k)&=-\eta_{\mu\nu} k^2\nn\\
g_m^{-1}(k)&=\fsl{k}-\hat m
\end{align}
where $\eta_{\mu\nu}$ is the metric tensor, and $\hat m=m-e\fsl{a}$ is a modified mass matrix term. 
Assuming that the $U(1)$ symmetry is not broken, in the physical vacuum $a^\mu=0$ and the mass term is $\hat m=m$.
The three vertices that come out from the interaction
are reported in the first line of Fig.4.

\begin{figure}[t] \label{fig:vertex}
\centering
\includegraphics[width=0.20\textwidth,angle=-90]{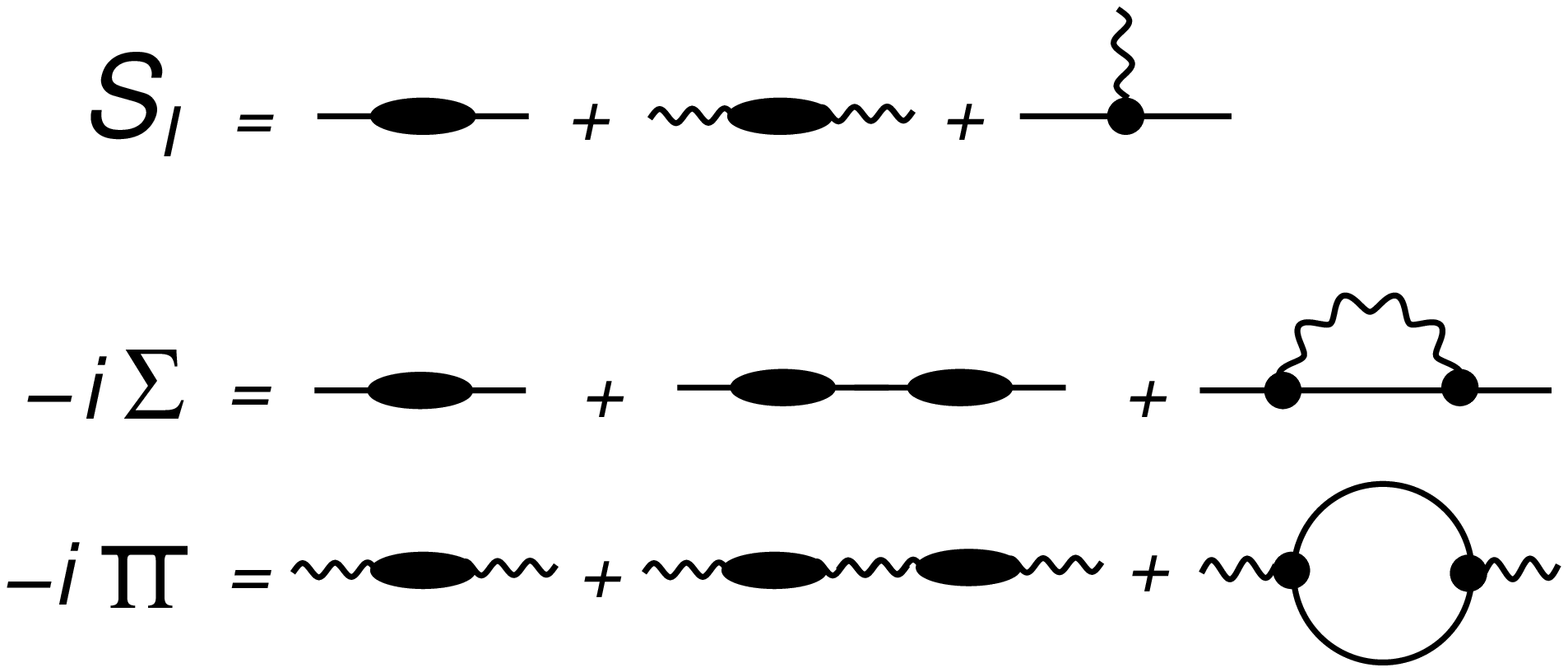}
\caption{The three vertices in the interaction $S_I$ of Eq.(\ref{SI2}) are shown in the first
line. First- and second-order graphs for the self-energy and polarization function 
are shown in the second and third lines, respectively. For each two-point function, we recognize 
a first-order graph, a reducible second-order graph, and a one-loop 1PI second-order graph.}
\end{figure}

The stationary conditions for the effective potential $V[a]=-\Gamma[a]/\Omega$ can be evaluated
by a straightforward extension of the connection to self-energy Eq.(\ref{delVns}).
Taking into account the matrix structure of the equations, for gauge fields the functional derivative
yields
\BE
\frac{\delta V_n}{\delta D_{\mu\nu} (k)}=\frac{i}{2} \left( \Pi^{\nu\mu}_n (k)-\Pi^{\nu\mu}_{n-1}(k)\right).
\label{delVnP}
\EE
Here, the self-energy is replaced by the polarization function $\Pi^{\mu\nu}$, which is the sum 
of connected two-point graphs, as shown in the third line of Fig.4 
where first- and second-order graphs are reported.
For  fermions,  we must add a minus sign because  self-energy graphs have a  loop less than 
the corresponding vacuum graphs  (a loop is removed by the functional derivative). Moreover,
we must drop the 2 factor that was inserted because of the permutation symmetry of two-point graphs 
for real fields. Taking into account the matrix structure, the connection to self-energy
reads
\BE
\frac{\delta V_n}{\delta G^{ab} (k)}=-i \left( \Sigma^{ba}_n (k)-\Sigma^{ba}_{n-1}(k)\right),
\label{delVnS}
\EE
where we inserted explicit spinor indices in the trial function $G^{ab}$ and in the  self-energy 
$\Sigma^{ab}$. First- and second-order self-energy graphs are shown in the second line of Fig.4.
Making use of Eqs.(\ref{delVnP}) and (\ref{delVnS}), the stationary conditions can be obtained 
without having to write the effective potential.

It is instructive to see what happens at first order; the variation of the trial functions $G$ and $D$
yields a set of two stationary conditions:
\begin{align}
\frac{\delta V^{(1)}}{\delta D_{\mu\nu} (k)} &=\frac{i}{2}  \Pi^{\nu\mu}_1 (k)=0  \nn\\
\frac{\delta V^{(1)}}{\delta G^{ab} (k)}&=-i  \Sigma^{ba}_1 (k)=0.
\end{align}
First-order self-energy and polarization are given by a single tree graph each, as shown in Fig.4.
The stationary conditions are equivalent to their vanishing
\begin{align}
-i \Pi^{\nu\mu}_1 (k)&=i\left[\Delta^{-1}_{\nu\mu}-D^{-1}_{\nu\mu}\right]=0  \nn\\
-i \Sigma^{ba}_1 (k)&=i\left[g_m^{-1}- G^{-1} \right]=0  \nn\\
\label{first}
\end{align}
yielding the trivial result $D=\Delta$ and $G=g_m$. Thus the GEP is equivalent to the free theory,
and any meaningful variational approximation requires the inclusion of second-order terms at least.

The failure of the first-order approximation would suggest that we look at the method of minimal variance, 
which is  a genuine second-order variational method. 
The proper self-energy and polarization contain one second-order term each, 
the one-loop graphs of Fig.4
\begin{align}
\Sigma_2^\star (k)&=i e^2 \int \ppp \gamma^\mu G(k+p)\gamma^\nu D_{\mu\nu}(p) \nn\\
{\Pi_2^\star} ^{\mu\nu}(k)&=-i e^2 \int \ppp
\Tr\left\{ G(p+k) \gamma^\mu  G(p)\gamma^\nu\right\} 
\label{proper}
\end{align}
These would be the usual proper two-point functions of QED if the  functions $D$ and 
$G$ were replaced by the bare propagators  $\Delta$ and $g_m$. 
The total second-order contributions to the two-point functions follow by the sum of 
all second-order graphs in Fig.4
\begin{align}
\Sigma_2 &=\left[g_m^{-1}-G^{-1}\right]\cdot G\cdot \left[g_m^{-1}-G^{-1}\right]
+\Sigma_2^\star\nn\\
\Pi_2 &=\left[\Delta^{-1}-D^{-1}\right]\cdot D\cdot \left[\Delta^{-1}-D^{-1}\right]
+\Pi_2^\star
\label{second}
\end{align}
where matrix products have been introduced in the notation. 
According to  Eqs.(\ref{delVnP}) and (\ref{delVnS}),  
the stationary conditions for minimal variance can be written as  
\begin{align}
\frac{\delta V_2}{\delta D_{\mu\nu} (k)}&=\frac{i}{2} \left( \Pi^{\nu\mu}_2 (k)-\Pi^{\nu\mu}_{1}(k)\right)=0\nn\\
\frac{\delta V_2}{\delta G^{ab} (k)}&=-i \left( \Sigma^{ba}_2 (k)-\Sigma^{ba}_{1}(k)\right)=0.
\end{align}
By insertion of the explicit expressions for first-order functions, as given by Eq.(\ref{first}), and
second-order functions Eq.(\ref{second}), the coupled equations can be recast as
\begin{align}
G(k)&=g_m(k)-g_m(k)\cdot \Sigma_2^\star (k)\cdot g_m (k)\nn\\
D_{\mu\nu}(k)&=\Delta_{\mu\nu}(k)-\Delta_{\mu\lambda}(k)\cdot 
{\Pi_2^\star}^{\lambda\rho}(k)\cdot \Delta_{\rho\nu}(k)
\label{stationary}
\end{align}
where the proper functions $\Pi_2^\star$, $\Sigma_2^\star$ are given by Eq.(\ref{proper}).
While this result resembles the simple lowest order approximation 
for the propagators in perturbation theory, it differs from it in two important ways: 
the presence of a minus sign in front of the second-order term, 
and the functional dependence on the unknown propagators $D$, $G$ in the
proper functions in Eq.(\ref{proper}). Because of this dependence, the stationary conditions are
a set of coupled integral equations, and their self-consistent solution is equivalent to the sum of
an infinite set of Feynman graphs. In fact, despite the appearance, the stationary conditions are not
a second-order approximation of an expansion in powers of the coupling $e^2$, but they make sense
even when the coupling is large as they derive from a variational constraint on the variance.

It is instructive to have a look at the second-order propagator $G^{(2)}$ 
as obtained by standard perturbation theory with optimized interaction $S_I$ 
and with free-particle propagators $G$, $D$  defined by 
Eq.(\ref{stationary}). 
We assume that the $U(1)$ symmetry is not broken, and $a=0$ in the physical vacuum.
In terms of the proper self-energy,
\BE
G^{(2)} (k)=\left[ G^{-1}(k)-\Sigma_1(k)-\Sigma_2^\star(k)\right]^{-1}
\EE
and by inserting the explicit expressions for the first-order self-energy $\Sigma_1=G^{-1}-g_m^{-1}$ 
and the bare propagator $g_m$, we find
\BE
[G^{(2)} (k)]^{-1}=\fsl{k}-m -\Sigma_2^\star(k)
\label{G2b}
\EE
which looks like the standard one-loop result of QED but differs for the functions $G$ and $D$  
that must be inserted in the one-loop $\Sigma_2^\star$ in Eq.(\ref{proper}) instead of the bare
propagators $g_m$, $\Delta$. If we expand the stationary conditions Eq.(\ref{stationary}) in
powers of the coupling $e^2$, take the lowest order approximation $G\approx g_m$, 
$D\approx\Delta$, and substitute back in the one-loop proper self-energy $\Sigma_2^\star$, then
Eq.(\ref{G2b}) becomes exactly equal to the one-loop propagator of QED. Thus the present variational
method agrees with the standard results of perturbation theory when the equations are expanded
in powers of the coupling.

As  a weaker approximation, the variational method can be set by solving one only of
the two stationary equations, while keeping one of the trial functions 
at a fixed value, out of the stationary point.
For instance, we could keep $D$ fixed at its free-particle value $D=\Delta$ and search for the
minimal variance by a variation of the trial function $G$. That is equivalent to looking for a solution
to the first part of Eq.(\ref{stationary}), which becomes a linear integral equation for $G$.
A path like that  leads to a Volterra integral equation\cite{varqed} that has 
a unique solution and can be solved by numerical iterative techniques. While that is proof of the
existence of the solution, a full numerical study of the set of coupled equation (\ref{stationary})
would be interesting for its eventual extension to non-Abelian gauge theories with large couplings. 
Of course, a regularization of diverging integrals and renormalization of bare couplings would be
required before attempting any numerical study. That is not a major problem\cite{varqed}  
and can be addressed by a  perturbative technique with the optimized interaction
$S_I$ that plays the role of the perturbation, and a set of  renormalization constants that can
be evaluated order by order, as shown for the  PGEP in Ref.\cite{stancu}.
Details of renormalization and a deeper study of the stationary conditions Eq.(\ref{stationary}) are
out of the aim of the present work and will be the content of another paper\cite{varqed}.

\section{Discussion and conclusion}

Let us summarize the main findings of the paper. 
While we are aware that many important aspects have not been addressed, 
like renormalization, gauge invariance, numerical study of the stationary equations, etc., 
the content of this paper is just a step towards a consistent development of 
variational methods for a better understanding of nonperturbative sectors of the standard model. 
While lattice simulations are the standard reference for nonperturbative calculations, 
an alternative analytical approach would be very valuable and welcome. Unfortunately, the GEP  is not
suitable for gauge theories, and even its second-order extension by the PGEP  seems to be useless\cite{stancu2}. 

We have shown that a viable general extension of the variational methods can be obtained
by using a trial function instead of a fixed shape for the free propagator. The stationary condition
on the effective potential becomes a set of integral equations for the unknown trial propagators
that is equivalent to optimize an infinite set of variational parameters.
Moreover, the stationary conditions are derived by the self-energy, 
without having to find the full effective potential. 
That simplifies the derivation and has been proven to be possible at any order
because of an exact connection between self-energy and functional derivatives of the effective
potential. 

Some important consequences of the variational equations have been proven for the scalar theory,
where the pole of the propagator is shown to be given by the simple first-order gap equation 
that seems to be more robust than expected by the PGEP analysis.
The second-order extension does not change the Gaussian gap equation but changes the shape 
of the trial function from its first-order free-particle form. 
That would also explain why no solution is found for the stationary
condition in the PGEP of Ref.\cite{stancu}, where a free-particle trial function is used:
the lack of any solution could be the sign of having chosen a wrong trial function.

Among the different variational strategies, the method of minimal variance has been shown 
to be suitable for gauge theories, where first-order approximations are useless. 
The method has been tested on QED and nontrivial results have been found. 
While the variational equations should hold for any strong value of the coupling, 
we have shown that an expansion in powers of $e^2$ gives back the standard results of QED. 
Of course, the aim is the extension to non-Abelian gauge theories 
in the strong coupling limit. With some constraint, the variational equations have been proven 
to admit a unique solution\cite{varqed} since the
equations can be recast in the form of  Volterra integral equations, and can be solved by iteration.
Some further numerical work is required for a deeper understanding of these findings, and
their eventual extension to important nonperturbative sectors of the standard model.

\appendix

\section{Explicit evaluation of the effective potential up to second order}

For the scalar theory of Section IIB, the connected vacuum 1PI graphs are displayed in 
Fig.1 up to second order. The sum of terms up to first order
is given in Eq.(\ref{V1}). There are five second-order graphs. The first graph 
in the second line of Fig.1 yields 
\BE
i\Gamma_2^{a}=\frac{\Omega}{4} \int\ppp g(p)[g_M^{-1}(p)-g^{-1}(p)]
g(p)[g_M^{-1}(p)-g^{-1}(p)]
\EE
and neglecting an additive constant term,
\BE
V_2^a=\frac{i}{2} \left[\frac{1}{2} \int\ppp g^2(p)g_M^{-2}(p)
-\int\ppp g(p)g_M^{-1}(p)\right].
\EE
The second graph in the second line of Fig.1 is
\BE
i\Gamma_2^{b}=6 v_4 \Omega I_0\int\ppp g(p)[g_M^{-1}(p)-g^{-1}(p)]g(p)
\EE
and yields
\BE
V_2^b=\frac{i}{2} \left[12iv_4 I_0^2+12 v_4 I_0\int\ppp g^2(p)g_M^{-1}(p)\right].
\EE
The third graph in the second line of Fig.1 is
\BE
i\Gamma_2^{c}=3i v^2_3 \int [g(x,y)]^3{\rm d}^4x \>{\rm d}^4y 
\EE
and then
\BE
V_2^c=\frac{i}{2} (i v^2_3) I^{(3)}.
\EE
The first graph in the third line of Fig.1 gives
\BE
i\Gamma_2^{d}=-12 v^2_4 \int [g(x,y)]^4{\rm d}^4x \>{\rm d}^4y 
\EE
yielding
\BE
V_2^d=\frac{i}{2} (i v^2_4) I^{(4)}.
\EE
The last graph of Fig.1 gives
\BE
i\Gamma_2^{e}=(6v_4I_0)^2 \int [g(x,y)]^2{\rm d}^4x \>{\rm d}^4y 
\EE
and then
\BE
V_2^e=\frac{i}{2}\left[i(6v_4I_0)^2 I^{(2)}\right].
\EE
The second-order contribution to the effective potential is the sum
\BE
V_2=V_2^a+V_2^b+V_2^c+V_2^d+V_2^e.
\EE
It can be easily checked that for the special case $g^{-1}(k)=k^2-M^2_2$, the 
total second-order potential $V_2$ becomes equal to its expression in Ref.\cite{stancu}.

The first-order self-energy is the sum of the two graphs
in the first line of Fig.2,
\BE
\Sigma_1 (k)=g^{-1}(k)-g^{-1}_M(k)-12 v_4I_0.
\EE
The second-order contribution to the self-energy is the sum of the eight 
second-order graphs in Fig.2. In the second line of Fig.2 we find four 1PI graphs: 
the first graph is
\BE
\Sigma_2^b(k)=-12 v_4 I_0+12i v_4 \int\ppp g^2(p)g_M^{-1}(p),
\EE
the second graph is
\BE
\Sigma_2^e(k)=-2(6v_4)^2 I_0  I^{(2)},
\EE
the third graph gives
\BE
\Sigma_2^c(k)=v^2_3 J_3(k),
\EE
and the fourth graph gives
\BE
\Sigma_2^d(k)=v^2_4 J_4(k).
\EE
The four reducible graphs in the third line of Fig.2 give the following
contributions: the first graph can be written as
\BE
\Sigma_2^a(k)=\left[g^{-1}(k)-g^{-1}_M(k)\right]+
\left[g(k)g^{-2}_M(k)-g^{-1}_M(k)\right],
\EE
the second graph is
\BE
\Sigma_2^f(k)=4(6v_4I_0)^2g(k)
\EE
and the sum of the last two graphs can be written as
\BE
\Sigma_2^{g}(k)=4! v_4 I_0 g(k)g^{-1}_M(k)-4!v_4I_0.
\EE

The functional derivative $\delta V_2/\delta g$ follows term by term:
\begin{align}
-2i\frac{\delta V_2^a}{\delta g(k)}&=g(k)g_M^{-2}(k)-g_M^{-1}(k)=\nn\\
&=\Sigma_2^a(k)-\Sigma_1(k)-12v_4 I_0,
\end{align}
\begin{align}
-2i\frac{\delta V_2^b}{\delta g(k)}&=-24v_4 I_0+12i v_4 \int\ppp g^2(p)g_M^{-1}(p)\nn\\
&+24 v_4 I_0 g(k) g^{-1}_M(k)=\nn\\
&=\Sigma_2^b(k)+\Sigma_2^g(k)+12 v_4 I_0,
\end{align}
and making use of Eq.(\ref{Jn})
\begin{align}
-2i\frac{\delta V_2^c}{\delta g(k)}&=v_3^2 J_3(k)=\Sigma_2^c(k)\nn\\
-2i\frac{\delta V_2^d}{\delta g(k)}&=v_4^2 J_4(k)=\Sigma_2^d(k).
\end{align}
Finally, the last term gives
\begin{align}
-2i\frac{\delta V_2^e}{\delta g(k)}&=-2(6v_4)^2 I_0 I^{(2)}+4(6v_4 I_0)^2g(k)=\nn\\
&=\Sigma_2^e(k)+\Sigma_2^f(k).
\end{align}
Putting together all these terms, the functional derivative of $V_2$ reads
\BE
-2i\frac{\delta V_2}{\delta g(k)}=\Sigma_2(k)-\Sigma_1(k)
\label{sigmaA}
\EE
where the second-order self-energy term is
\BE
\Sigma_2=\Sigma_2^a+\Sigma_2^b+\Sigma_2^c+\Sigma_2^d+\Sigma_2^e+\Sigma_2^f+\Sigma_2^g,
\EE
in perfect agreement with Eqs.(\ref{delVns}) and (\ref{delV2aa}).


\begin{thebibliography} {99}

\bibitem{kogan} I.L. Kogan and A. Kovner, Phys. Rev. D {\bf 52}, 3719 (1995) 
\bibitem{reinhardt} C. Feuchter and H. Reinhardt, Phys. Rev. D {\bf 70}, 105021
(2004).
\bibitem{szcz} A. P. Szczepaniak, Phys.Rev. D {\bf 69}, 074031  (2004). 
\bibitem{schiff} L.I. Schiff, Phys. Rev. {\bf 130}, 458 (1963).
\bibitem{rosen}G. Rosen, Phys. Rev. {\bf 172}, 1632 (1968).
\bibitem{barnes}T. Barnes and G. I. Ghandour, Phys. Rev. D {\bf  22 }, 924 (1980).
\bibitem{stevenson} P.M. Stevenson, Phys. Rev. D {\bf 32}, 1389 (1985).
\bibitem{ibanez} R. Iba\~nez-Meier, I. Stancu, P.M. Stevenson, Z. Phys. C {\bf 70}, 307 (1996).
\bibitem{var} F. Siringo,  Phys. Rev. D {\bf 62}, 116009 (2000).
\bibitem{light} F. Siringo, Europhys. Lett. {\bf 59}, 820 (2002).
\bibitem{bubble} F. Siringo and L. Marotta, Int. J. Mod. Phys. {\bf A25}, 5865 (2010), arXiv:0901.2418v2.
\bibitem{superc1} M. Camarda, G.G.N. Angilella, R. Pucci, F. Siringo, 
Eur. Phys. J. B {\bf 33}, 273 (2003).
\bibitem{superc2} L. Marotta, M. Camarda, G.G.N. Angilella and F. Siringo, 
Phys. Rev. B {\bf 73}, 104517 (2006).
\bibitem{kim}  C. K. Kim, A, Rakhimow, Jae Hyung Hee, Eur. Phys. Jour. B {\bf 39}, 301 (2004).              
\bibitem{su2} F. Siringo, L. Marotta, Phys. Rev. D {\bf 78}, 016003 (2008).
\bibitem{LR} F. Siringo and L. Marotta, Phys. Rev. D {\bf 74}, 115001 (2006).
\bibitem{AF} L. Marotta and F. Siringo, Mod. Phys. Lett. B, {\bf 26}, 1250130 (2012), arXiv:0806.4569v3.
\bibitem{HT} F. Siringo, Phys. Rev. D {\bf 86}, 076016 (2012), arXiv: 1208.3592v2.
\bibitem{stancu} I. Stancu and P. M. Stevenson, Phys. Rev. D {\bf 42}, 2710 (1990).
\bibitem{stancu2} I. Stancu, Phys. Rev. D {\bf 43}, 1283 (1991).
\bibitem{weinbergII} S. Weinberg, {\it The quantum theory of fields}, Vol.II, Cambridge University Press (1996). 
\bibitem{tedesco} P. Cea, L. Tedesco, Phys. Rev. D {\bf 55}, 4967 (1997).
\bibitem{minimal} P. M. Stevenson, Phys. Rev. D {\bf 23}, 2916 (1981).
\bibitem{sigma} F. Siringo and L. Marotta, Eur. Phys. J. C {\bf  44}, 293 (2005).
\bibitem{sigma2} F. Siringo, arXiv:1308.4037.
\bibitem{varqed} F. Siringo, arXiv:1308.2913.
\end{thebibliography}
\end{document}